\documentclass{aa}	
\usepackage{graphicx,rotating,amsmath,epsf,txfonts,natbib}
\def\kms{\rm km~\;s$^{-1}$}
\def\etal{{et al. }}

\def\3o{O~{\sc iii}}
\def\4o{O~{\sc iv}}

\def\arcsec{$^{\prime\prime}$}

\begin{document}

\title{Coronal hole boundaries evolution at small scales: II. XRT view}
\subtitle{Can small-scale outflows at CHBs be a  source of the slow solar wind?  }
\author{S. Subramanian,  M. S. Madjarska \and J. G. Doyle}
\offprints{M.S. Madjarska, madj@arm.ac.uk}
\institute{Armagh Observatory, College Hill, Armagh BT61 9DG, N. Ireland}
           
\date{Received date, accepted date}
\abstract
{}
{We aim to further explore the small-scale evolution of coronal hole boundaries  
using X-ray high-resolution and high-cadence images. We intend to determine the 
fine structure and dynamics of the events causing the changes of the coronal hole 
boundaries and to explore the possibility that these events are the source of the 
slow solar wind. }
{We developed an automated procedure for the identification of transient 
brightenings in images from the X-ray telescope on-board Hinode taken with an Al 
Poly filter in the equatorial coronal holes, polar coronal holes,  and the quiet Sun with and without transient coronal holes. }
{We found that in comparison to the quiet Sun, the boundaries of  coronal holes 
are abundant with brightening events including areas inside  the coronal holes where 
closed magnetic field structures are present. The visual  analysis of these  brightenings 
revealed that around 70\% of them in equatorial, polar and transient coronal holes and  their boundaries show expanding loop structures and/or collimated outflows. In the quiet Sun only 30\% of the brightenings show flows with most of them appearing to be contained in the solar corona by closed magnetic field lines. This strongly suggests that magnetic reconnection of co-spatial  open and closed magnetic field lines  creates the necessary conditions for  plasma outflows  to large distances. The ejected plasma always originates from  pre-existing or newly emerging (at X-ray temperatures) bright points.}
{The present study confirms our findings that the  evolution of loop structures known as coronal bright points is associated with the small-scale changes of  coronal hole boundaries. The loop structures show an expansion and eruption with the trapped plasma consequently escaping along the ``quasi'' open magnetic field lines. These ejections appear to be triggered by magnetic reconnection, e.g. the so-called interchange reconnection between the closed magnetic field lines (BPs) and the open magnetic field lines of the coronal holes. We suggest that these plasma outflows are possibly one of the sources of the slow solar wind.}


\authorrunning{Subramanian \etal}
\titlerunning{Coronal hole boundaries evolution}

\maketitle
\section{Introduction}

Coronal holes (CHs) are regions of predominantly unipolar coronal magnetic 
fields with a significant component of the magnetic field open into the heliosphere. 
They are visible in spectral lines emitting at coronal temperatures as dark areas when 
compared to the quiet Sun, while in the chromospheric He~{\sc i}~10830~\AA\ line they 
appear bright. For detailed introduction on coronal holes see \citet{2009arXiv0906.2556M} (hereafter paper I). 
CHs are identified as the source of the fast solar wind 
with velocities of up to $\approx$~800~\kms\  \citep{1973SoPh...29..505K}.  In 
contrast, the slow wind has velocities around 400~\kms\ and  is more dense and 
variable in nature when compared to the fast solar wind. \citet{1996ASPC..109..491V} 
found from Ulysses satellite data  that the elemental composition of the fast 
wind is similar to the elemental composition of the photosphere. The slow solar  
wind is enriched with low first ionization potential (FIP) elements by a factor 
of 3--5 greater than in the photosphere (with respect to hydrogen) while higher 
FIP elements were found at solar surface abundances. The FIP effect describes 
the element abundance anomalies  (the enhancement of elements with low FIP such 
as Fe, Mg and Si over those with high FIP like Ne and Ar) in the upper solar 
atmosphere and solar wind, and can give a clue on the origin of both the fast 
and the slow solar winds. \citet{1996ASPC..109..491V} concluded that the fast and 
slow solar winds not only differ in their kinetics but also in their composition 
of elements. 

\citet{2004ApJ...612.1171W} suggested that the release of trapped plasma in closed 
loop structures by magnetic reconnection could play a significant role in the solar wind flow. Such reconnection between the open and  closed magnetic field 
lines presumably happens continuously at coronal hole boundaries. \citet{1998ApJ...498L.165W} investigated the ejection of plasma blobs from the streamer belt linked to 
the slow wind and concluded that magnetic reconnection between the distended 
streamer loops and the open magnetic field lines might be behind the plasma 
ejection. They also suggested that this ejection cannot account for all the 
slow solar wind and a major component should, therefore, originate outside the helmet 
streamers, i.e. from inside the coronal holes. \citet{2004ApJ...603L..57M} 
found non-Gaussian profiles  along the boundaries in an equatorial extension 
of a polar  CH in  the mid- and high-transition region lines N~{\sc iv}~765~\AA\ 
and Ne~{\sc viii}~770~\AA, respectively, recorded with the Solar Measurement of Emitted 
Radiation (SUMER) spectrometer on-board the Solar and Heliospheric Observatory (SoHO). 
The authors suggested that these profiles are the signature of magnetic 
reconnection occurring between the closed magnetic field lines of the quiet Sun 
and the open of the coronal hole. Similar activity was reported by 
\citet{2006A&A...446..327D} along the boundary of a polar CH.

\begin{figure*}[!ht]
    \centering
    \vspace{16cm}
      \includegraphics{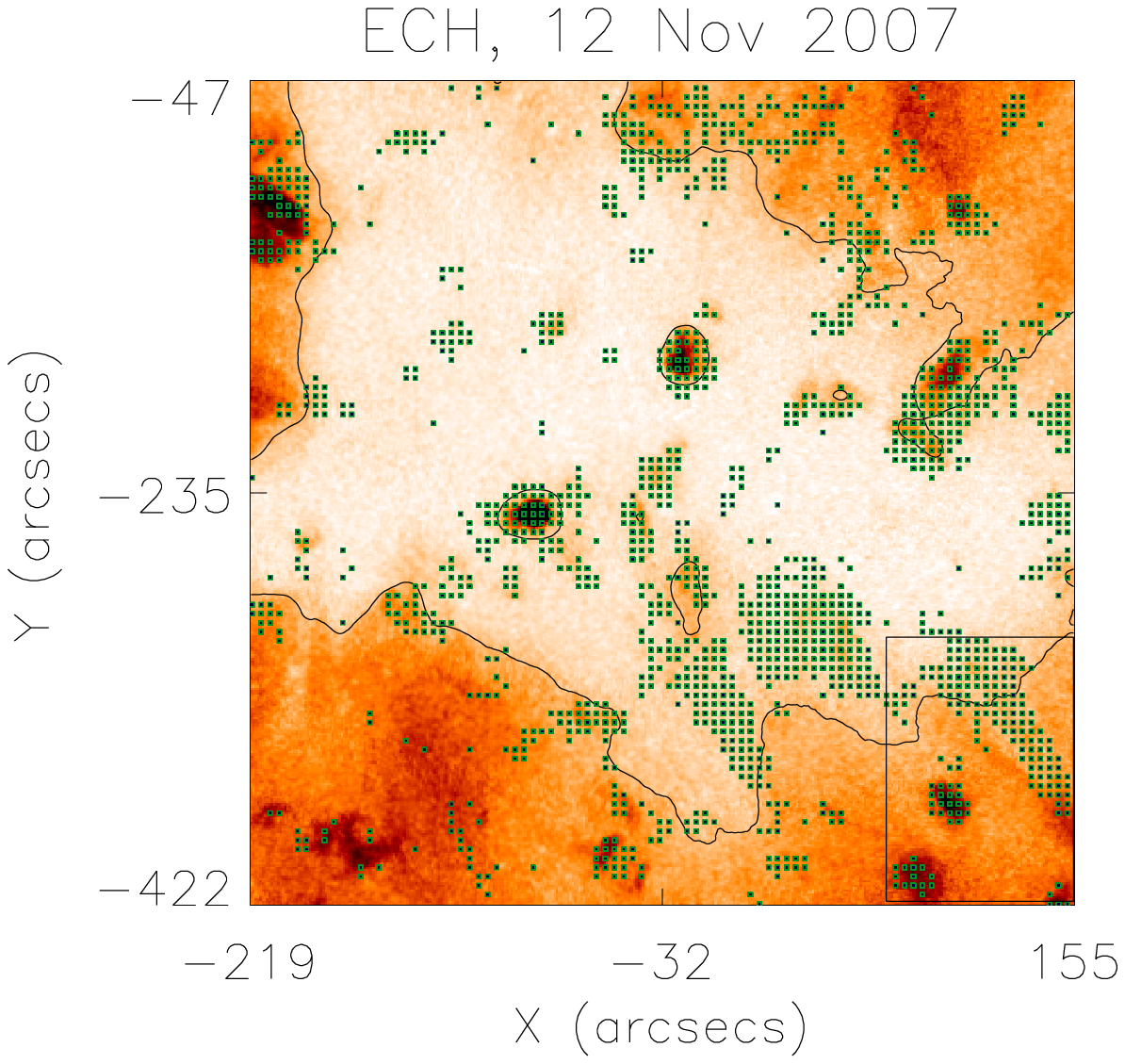}
      \includegraphics{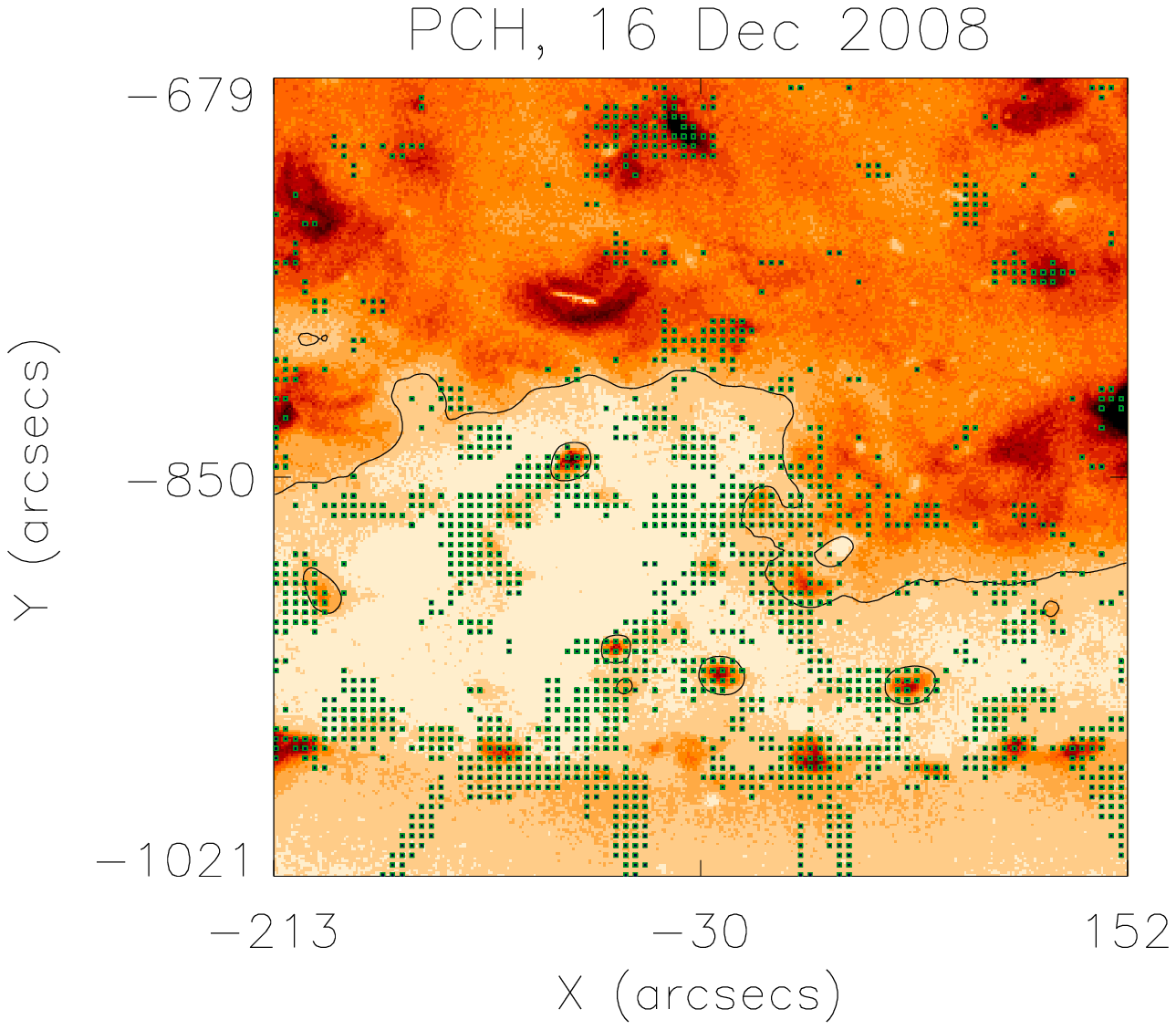}  
        \includegraphics{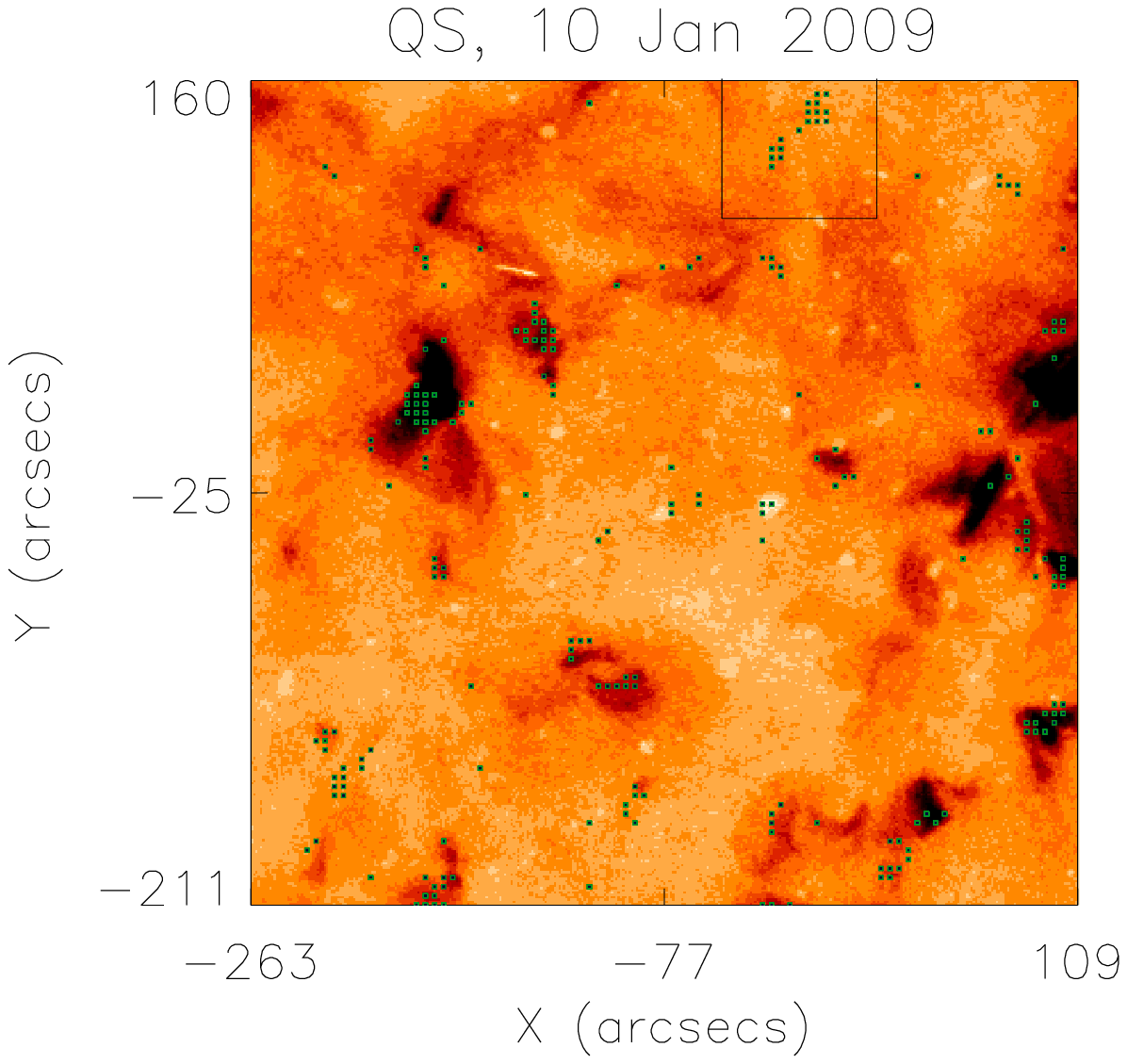}
          \includegraphics{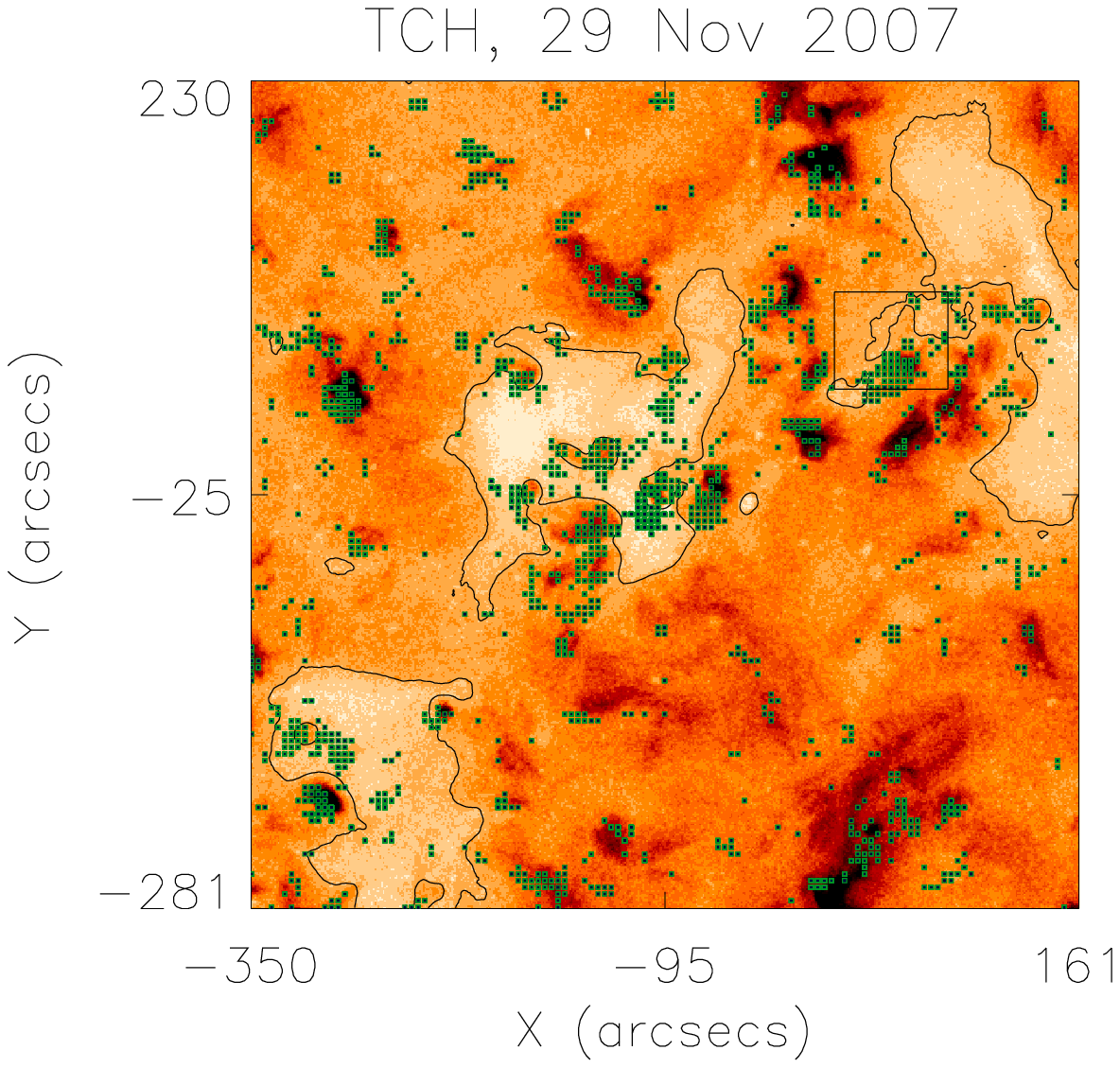}
\vspace*{-0.7cm}
      \caption{ Equatorial coronal hole (top left),  polar CH (top right), 
      quiet Sun (bottom left) and quiet Sun with TCHs (bottom right)  
      with the positions of all the corresponding identified brightening 
      pixels over-plotted. The CH boundaries are outlined with a black line. The 
      over-drawn rectangles correspond to the field-of-views shown in 
      Figs.~\ref{fig5}, \ref{fig6} and \ref{fig7}. }
      \label{fig1}
\end{figure*}


\begin{table*}
\begin{center}
\caption{Description of the  XRT  data used in the present study (ECH -- equatorial 
coronal hole, PCH -- polar coronal hole, TCH -- transient coronal hole). }
\label{table1}
\begin{tabular} {ccccc}
\hline
Date & Remark & Observing period & Field-of-view   & Exposure time 	\\
     & 	      &	of time (UT)	            & (arcsec) &  (sec)		\\
\hline
09/11/07 & ECH & 06:35-14:59  & $366\times366$	 & 16	\\
12/11/07 & ECH & 01:17-10:59  & $374\times374$	 & 16	\\
14/11/07 & ECH & 00:12-11:11  & $374\times374$	 & 16	\\	
16/11/07 & ECH & 18:07-23:58  & $370\times370$	 & 16	\\
16/12/08 & PCH	 & 10:02-17:52  & $362\times337$	 & 23	\\
20/09/07 & PCH 	 & 12:22-18:05  & $1018\times291$ 	 & 16	\\
10/01/09 & QS	 & 11:30-17:27  & $370\times370$	 & 23	\\
13/01/09 & QS& 11:22-17:41  & $370\times370$	 & 23	\\
29/11/07 & QS with TCH& 18:17-23:59  & $506\times506$	 & 23	\\
\hline
\end{tabular}
\end{center}
\end{table*}

In paper I we demonstrated that although isolated  equatorial  CH and 
equatorial extension of polar CH maintain their general shape during several 
solar rotations, a closer look at their day-by-day and even hour-by-hour 
evolution demonstrates significant dynamics. We showed that small-scale 
loops which are abundant along coronal hole boundaries contribute to the 
small-scale evolution of coronal holes. We suggested that these dynamics 
are triggered by continuous magnetic reconnection already proposed by
\citet{2004ApJ...603L..57M}. The next step of our research was to analyse  
images taken with the X-ray Telescope (XRT) on-board Hinode.
 
Seen in XRT images, CHs are highly structured and dynamic at small scales.  
High cadence  XRT data reveal in great detail the fine structure of coronal 
bright points (BPs) and  X-ray jets associated with them.  X-ray jets 
are collimated transient ejection of coronal plasma, first reported with the 
Solar X-ray telescope (SXT) onboard Yohkoh \citep{1992PASJ...44L.173S}. They 
are believed to result from magnetic reconnection \citep{1994xspy.conf...29S} and represent 
plasma outflows from the reconnection site. Recently,  \citet{2008ApJ...673L.211M}  presented  
three--dimensional simulations of flux emergence in CH combined with spectroscopic  and imager 
observations from XRT and EIS/Hinode of an X-ray jet.  The authors  report that a jet resulting from 
magnetic reconnection  is  expelled upward along the open
reconnected field lines with values of temperature, density, and velocity in agreement with the XRT and EIS
observations. \citet{1996PASJ...48..123S} reported 100 jets over 6 months in SXT 
images from the Yohkoh while \citet{2007PASJ...59S.771S} upgraded 
this number to an average of 60 jet events per day in polar coronal holes. The 
authors concluded that jets preferably occur inside polar coronal holes (PCH).  

We should note that although TRACE images  which have higher spatial 
resolution were used in paper I, the detailed structure of the dynamic 
changes along CH boundaries was hard to distinguish. A reason for that is 
the effect of  stray light in the TRACE  extreme-ultraviolet (EUV) telescope 
reported recently by \citet{2009ApJ...690.1264D}. The authors found that 43\% of the light 
which enters TRACE through the Fe~{\sc ix/x}  171~\AA\ filter is scattered
either through diffraction off the entrance filter grid or through other 
non-specific effects. This creates a haze effect and  especially effects  
the visibility of small-scale bright structures. 

\begin{figure}[!h]
\vspace{6cm}
\includegraphics{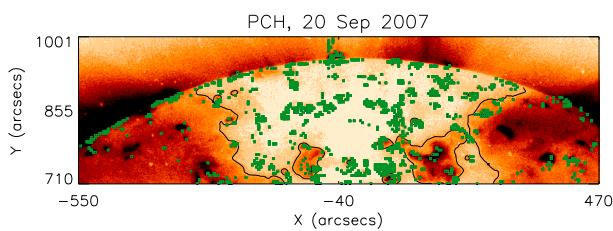}
\vspace{-2.9cm}
\caption{Polar coronal hole observed by XRT on 2007  September 20 with  the positions of all the identified brightening pixels over-plotted.}

\label{fig2}
\end{figure}


  \begin{figure}[!h]
  \vspace{7cm}
\includegraphics{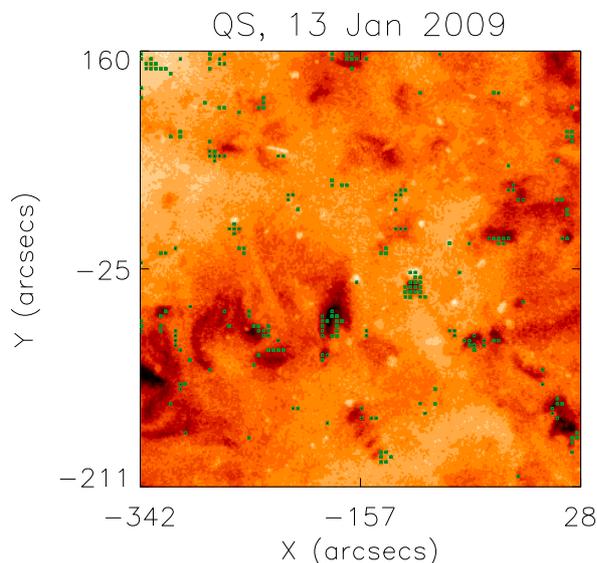}
      \caption{Quiet Sun observed by XRT on 2009  January  13 with  the positions 
      of all the identified brightening 
      pixels over-plotted.}
      \label{fig3}
\end{figure}

Other transient structures seen in coronal holes  are the so-called plumes 
observed off-limb above the North and South polar coronal holes.  They were 
first observed in white light as ray like structures 
\citep{1965PASJ...17....1S}. They are also observed at EUV 
and soft X-ray temperature \citep{1978SoPh...58..323A} as coronal outflow 
structures similar to coronal jets, but hazy in nature with no sharp boundaries unlike jets.
They represent denser and cooler outflows with respect to the surrounding media and are 
observed to extend from coronal BPs. They can extend up to 30 R$_{\odot}$  from the solar disk center in a plane 
image \citep{2001ApJ...546..569D} and are observed to be in a steady state 
for at least 24 hours \citep{1997SoPh..175..393D}. The X-ray jets have been 
identified as precursors for the plume formation  \citep{2008ApJ...682L.137R}. 
Recently, \citet{2008SoPh..249...17W} identified coronal plumes inside equatorial 
coronal holes. They found that the plumes are analogous to polar coronal 
plumes. On the disk they are seen as a diffuse structure with a bright core 
and associated with EUV BPs. 

The present study is a continuation of paper I and presents results from 
the analysis of high-cadence/high-resolution images of coronal holes (equatorial, 
polar and transient) and quiet Sun from XRT/Hinode. We aim to establish which type of 
event generates the non-Gaussian profiles registered at CH boundaries by  
\citet{2004ApJ...603L..57M} and how they are related to the small-scale BPs 
evolution along coronal hole boundaries as reported in paper I. In 
Sect.~\ref{data} we describe the data used for our study. Sect.~\ref{ip} 
outlines an automatic brightening identification  procedure. In 
Sect.~\ref{results}, we give the obtained results and draw some 
conclusions on the outcome of our study. Finally, in Sect.~\ref{conclusions} 
we discuss the implication of our result to the understanding of the nature 
of coronal hole boundaries evolution at small scale and the possible 
contribution of these events to the formation of the slow solar wind.  

\begin{figure*}[!ht]
    \centering
    \vspace{7cm}
      \includegraphics{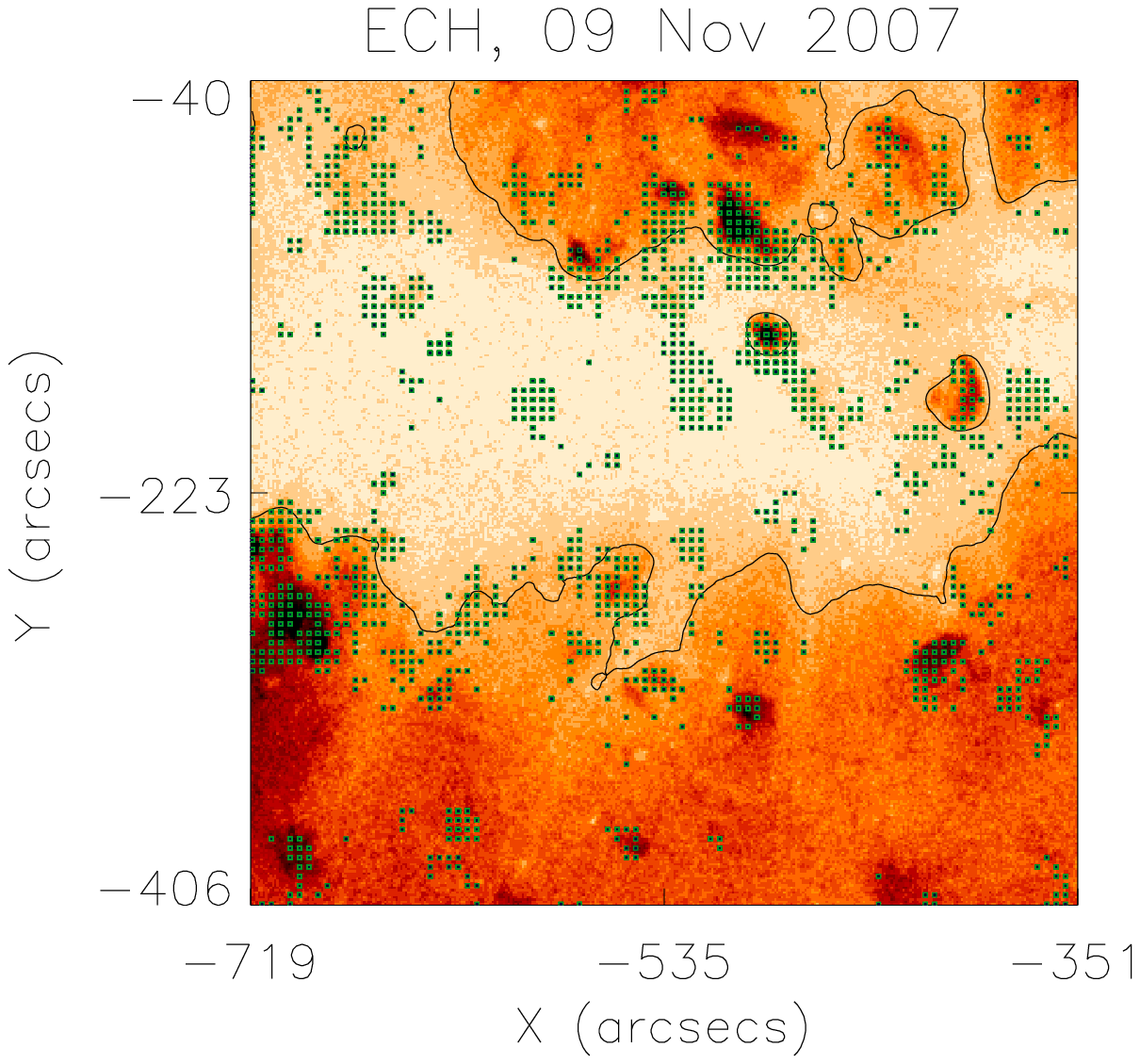}
      \includegraphics{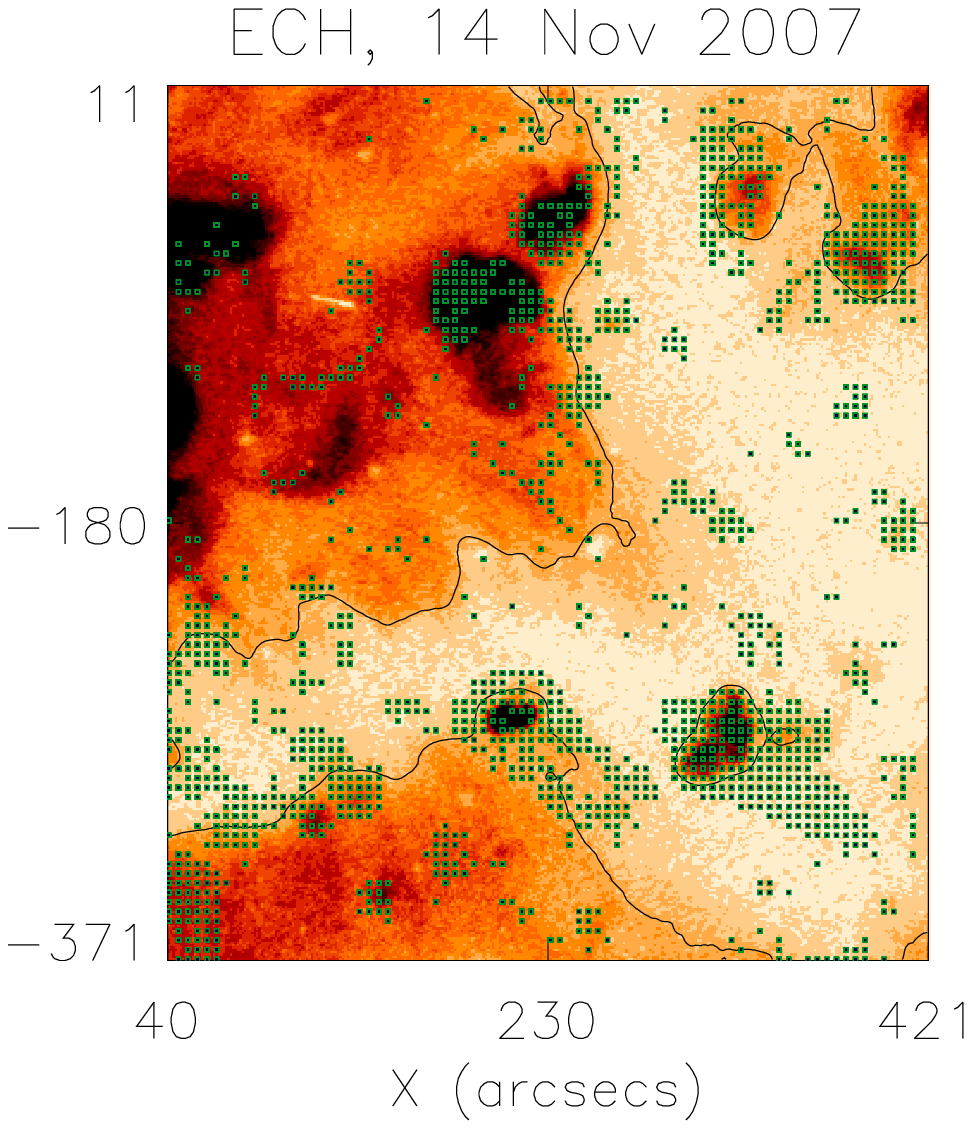}  
        \includegraphics{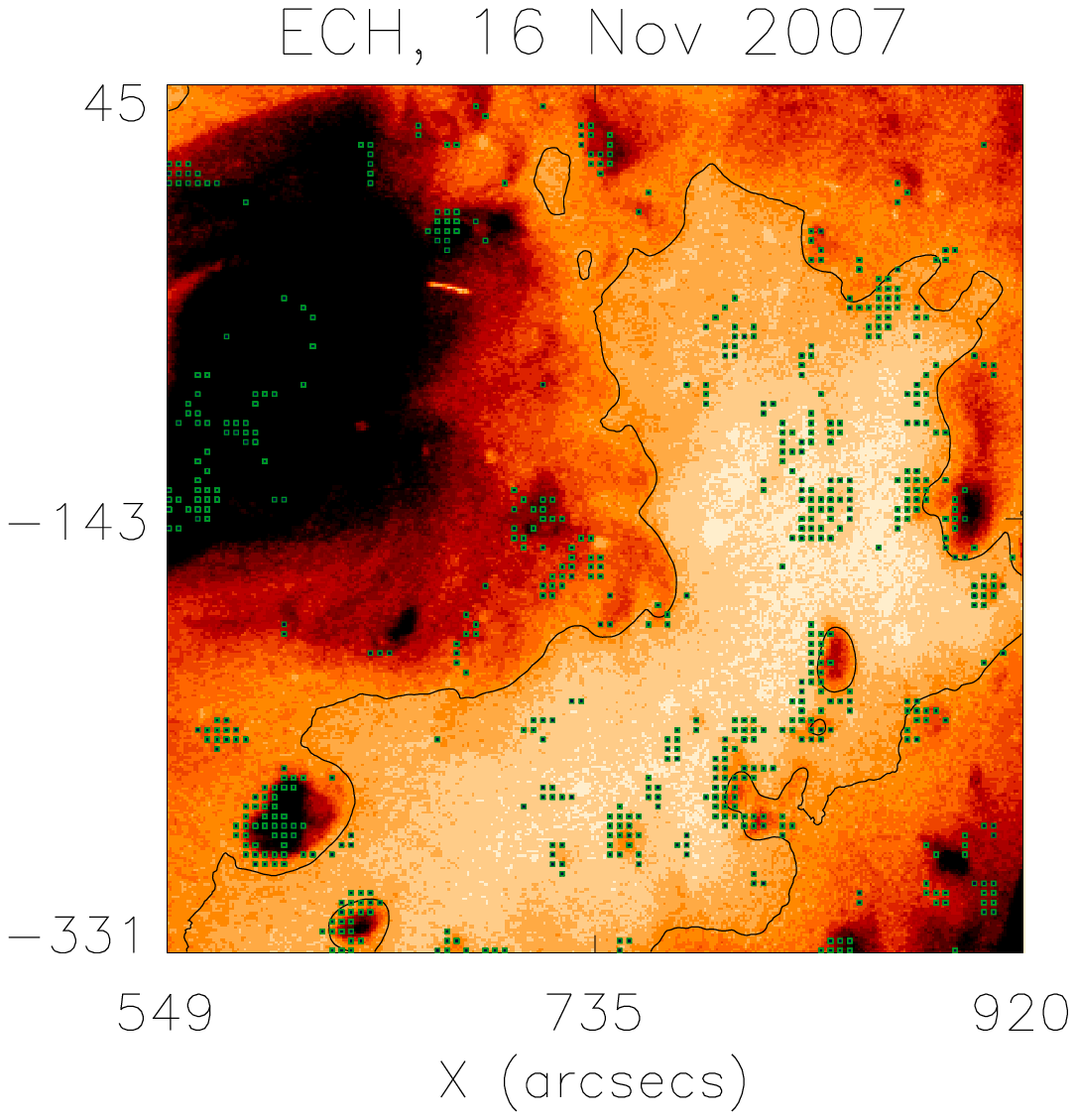}
\vspace*{-0.75cm}
     \caption{ Equatorial coronal hole observed by XRT on 2007  November  9, 14 and 16 with the positions of all the identified brightening pixels over-plotted. The CH boundaries  are over-plotted with a black solid line.}
      \label{fig4}
\end{figure*}
\section{Data, reduction and preparation}
\label{data}

We used images from the X-ray Telescope  \citep{2007SoPh..243...63G} on-board 
Hinode taken during a dedicated observing run of an isolated equatorial 
coronal hole (ECH), a Southern polar coronal hole and quiet Sun regions. 
The ECH was tracked from the West to the East limb from 8 to 10 hours per day 
for 4 days. The Southern polar CH was 
observed for one day while the quiet Sun regions over 2 days. All data were 
taken with an Al Poly filter which has a well pronounced temperature response 
at logT$_{max}~\approx$~6.9~K. XRT images have an angular pixel size of
1\arcsec~$\times$~1\arcsec\ at full resolution. They were recorded with 16~{\rm s} and 
23~{\rm s} exposure time and  a cadence of about 40~{\rm s}. We also used 
randomly selected  quiet Sun data with transient coronal holes (TCHs) and a Northern 
polar coronal hole observation. Further details on the data can be found 
in Table~\ref{table1}.

The data were reduced using the standard procedures, which include flat-field 
subtraction, dark current removal, despiking, normalisation to  data number 
per second to account for the variations in exposure time, satellite jitter 
and orbital variation corrections. The images were then de-rotated to a 
reference time to compensate for the solar rotation. A common field-of-view 
(FOV) was selected from all the images for each day. We then prepared an 
array with dimensions (nx, ny, nf), where nx is the number of Solar\_X pixels, 
ny is the number of Solar\_Y pixels and nf  is the number of images. Each image was 
binned to   4~$\times$~4 pixels$^2$ in order to improve the signal-to-noise ratio and 
 reduce the data points (and subsequently 
the computational time).  The binned images were used to produce light 
curves of nf points for each pixel. These light curves were the input for 
an identification procedure which will be discussed  in the next section.

\section{Brightening identification procedure}
\label{ip}

We developed an automatic identification procedure to distinguish small-scale 
intensity enhancements in XRT images. While the visual identification of large 
events such as jets from bright points give good results, it is difficult to 
identify and track small-scale events, especially on the quiet Sun high 
background emission or over pre-existing bright coronal loop structures 
(e.g. bright points, active regions etc). We eliminated all light curves which show no activity or minimum activity comparable to the noise level.  

\begin{table*}
\begin{center}
\caption{Number of events  over  24 hours per 100 $\times$ 100 arcsec$^2$ identified inside the  coronal holes (CH), in coronal hole boundary regions (CHBR)  and  in the quiet Sun (QS).}
\label{table2}
\begin{tabular} {lcccccc}
\hline \\
 & \multicolumn{3}{c}{No of events identified} & \multicolumn{3}{c}{No of unresolved events } \\
Date  & \multicolumn{3}{c}{with plasma outflows} & \multicolumn{3}{c}{identified with no plasma outflows} \\
\hline
 & CH & CHBR & QS & CH & CHBR & QS \\ 
\hline 
09/11/07 & 29  & 40   &    6    &   6  & 26  &  8      \\
12/11/07 & 16  & 57   &    6    &   2  & 17  &  7      \\
14/11/07 & 10  & 51   &    1    &  16  & 41  &  7      \\
16/11/07 & 56  & 32   &    -    &  44  & 33  & 12      \\
16/12/08 & 99  & 86   &    6    &  31  & 14  & 33      \\
20/09/07 & 57  & 72   &    4    &   5  &  7  &  9      \\
10/01/09 &  -  &  -   &    6    &   -  &  -  & 17      \\
13/01/09 &  -  &  -   &    9    &   -  &  -  & 14      \\
29/11/07 &  -  & 64   &    7    &   -  & 62  & 14      \\ 
\hline
\end{tabular}
\end{center}
\end{table*}

The first step of the identification procedure was to define the background 
emission for each light curve. The light curves were smoothed over a  window 
of width 5 to remove the spikiness in the background. Due to a difference in the 
background emission between the quiet Sun and the CHs, it was necessary to set two 
different thresholds for further analysis. The threshold we used was 
1.8 times the mean emission value  for the CHs and 1.3 times the  
mean emissivity for the QS. The comparatively higher threshold set for the 
CH light curves helped to eliminate the high fluctuations of the low emission 
background. Light curves with a maximum value less than these thresholds were 
neglected. Any point in the light curve was considered as a peak if its 
value was greater than the threshold and also greater than the average of the
two preceding points, and the average of two  successive points. All the 
values below the threshold were considered as local minima. Each identified 
peak was traced back on either side to identify the minimum from the local 
minima. The value of all the points between the two identified minima for each 
peak were set to zero in the light curve and thereby from the average over 
the rest of the light curve (I$_{av}$) we computed the standard deviation (SD). 
The new background (BG) was obtained as BG = I$_{av}$ +1.1~$\times$~SD.

The next step was the actual  identification of intensity enhancements. A  
new threshold of 2~$\times$~BG for CHs and 1.3~ $\times$~BG for QS was set 
using the above  calculated background. Intensity increases above these 
thresholds with  corresponding minima less than  BG and duration less than 45 
minutes were identified. A pixel brightening  was considered only if all the 
above mentioned conditions were satisfied. The threshold was calculated with 
a trial and error method.

The peaks having a duration of more than 45 minutes were examined separately. 
The closest local minimum on either side of the peak were traced back. If 
the difference between the peak and the minimum were greater than the BG for 
the CHs and 0.3$\times$BG for the QS with the duration less than 45 minutes, then 
they were considered. Also the peaks which have one minimum that was either 
in the beginning or at the end of the light-curve were evaluated with the same 
criteria, in order not to miss any real event. Any intensity enhancements 
in the coronal holes, the quiet Sun or over pre-existing bright loop structures 
 which satisfy the above criteria could be 
identified by our procedure. 

\section{Results and discussion}
\label{results}

As it has been described in Sect.~2, we made a selection of data which comprised 
observations of different features on the  Sun: equatorial  and polar coronal holes 
as well as  quiet Sun regions  with and without transient coronal holes.  
In Fig.~\ref{fig1}--\ref{fig4} we display examples of an X-ray image from each different region.  
Our intention was to find out whether the changes we have seen so far along CHBs 
(\citet{2004ApJ...603L..57M} and paper I) are unique for CH regions, i.e. regions 
of open magnetic field lines.  These data also permit to resolve the fine 
structure of individual features and follow their dynamics at high cadence. 
 To each dataset we applied the identification procedure described above. This 
procedure provided us with the following information:  (i) light curves which 
contain one or more  radiance enhancement identified as brightenings following 
the criteria given in Sect.~ 3; (ii)  the start  and end time of each radiance 
enhancement; (iii)  the brightening positions in pixel numbers. As we produced 
light curves by binning over 4~$\times$~4 pixels$^2$, imprints of brightening 
events with spatial scales larger than 4~$\times$~4 pixels$^2$ were observed 
in more than one light curve. This made visual grouping of identified bright 
pixels essential to distinguish each event.  Grouping of the features into 
individual events was done by playing the image sequence of each dataset with 
the identified brightenings over-plotted at corresponding times (see the online 
movies). Clusters of bright pixels identified next to each other with similar 
lightcurves were grouped into events. The events showing plasma outflows (i.e. 
plasma moving along quasi-straight trajectories) were classified as jets, while 
events exhibiting plasma blobs moving along curved trajectories  or just 
brightening increase in a group of pixels were classified as unresolved 
brightenings events. The so-called space-time plot was also used to 
investigate the plasma motion in the form of a jet and to determine their proper 
motion. A space-time plot was produced by averaging  over a slice of 3 pixels wide and 100 
pixels long from each image, cut along the jet, i.e. in the direction of plasma propagation and then plotting that in time 
\citep[][and Subramianian, PhD thesis 
2010]{2007PASJ...59S.745S} .  We were able to group more 
than $95\%$ of the identified bright pixels. The ungrouped pixels ($\leq5\%$) 
comprise  bright pixels identified at the edges of images and above bad 
pixels. The pixels identified in the beginning of each dataset which 
could not be classified due to the lack of coverage of the whole event and 
the pixels identified with a time lapse over their lifetimes were also rejected 
from counting. 

The visual grouping of identified bright pixels into 
events can be found in Table \ref{table2}. We defined a coronal hole boundary 
region (CHBR) as the region  $\pm$15\arcsec\ on both sides of the contour line 
defining the CH boundary.  Additionally, animated image sequences with over-plotted  
identified brightenings at  corresponding times can be seen online as 
movie\_qs.mp4 (quiet Sun region on 2009 January 10), movie\_ech.mp4 (ECH on 
2007 November 12,), movie\_pch.mp4 (PCH on 2008 December 16) and movie\_tch.mp4  
(TCH on 2007 November 29). 

The first and the most important result of this study is easily noticeable 
from Figs.~\ref{fig1}--\ref{fig4} the boundaries of coronal holes are abundant with 
brightening events which appear much larger than the same phenomena in the 
quiet-Sun region. We separated the events visually into two groups, events 
with plasma outflows or jet-like events  and events without outflows or 
simple brightenings. The equatorial coronal hole data, observed near the disk 
center, shows  twice  as many jet-like and simple brightening events in 
the CHBRs (as defined above) as compared to the CH regions. In contrast, polar 
coronal hole data and ECH close to the West limb (2007 November 16) show a higher 
number of events inside the CH as well as in the CHBRs suggesting that this  
can be due to the line-of-sight effect. However, further investigation is 
needed on a larger number of datasets.  
 
If we assume that the magnetic reconnection responsible for the occurrence 
of jet-like events takes place predominantly between closed and open 
magnetic field lines, then the number of reconnection events producing 
outflows will be always higher in the CH boundary region since open and 
closed magnetic field lines are continuously pushed together by different 
processes such as convection, differential rotation, meridian motions 
etc. Inside coronal holes, where the number of bipolar systems and the 
corresponding closed loop structures are limited, the number of jet-like 
events will therefore be lower. For the transient coronal hole regions, the 
separation of the coronal hole boundaries region  (30\arcsec\ wide) from the coronal holes, 
for estimating the number of events in each region, is more difficult due to the very small size of these coronal holes. Therefore,  here
 we consider that these CHs represent entirely a boundary region. The number of events found in these 
 TCHs is several times larger than in the quiet Sun (both with and without outflows).  

The plasma ejected during the outflow events always originates from 
pre-existing or newly emerging (at X-ray temperatures) bright points
both inside CHB regions and CHs. They typically start with a brightening 
in just a few XRT  pixels (4-6) somewhere in a pre-existing BP which 
we believe to be the reconnection site.  \cite{1996PASJ...48..123S} 
reported from the statistical study of 100 jets in SXT/Yohkoh observations 
that most of them were associated with micro-flares in the foot-points of 
the jets.  \citet{2007PASJ...59S.745S} resolved the fine structure of a 
quiet Sun X-ray jet to be the expansion and eruption of loop structures colliding 
with the ambient magnetic field, similar to the CH jet.  Madjarska (2009) studied in great detail 
one of the  jet-like phenomena identified in the datasets analysed here. The author estimated that the 
reconnection site reaches temperatures of up to 12 MK from observations 
of Fe~{\sc xxiii}~263~\AA\ from EIS/Hinode, which confirms that some of these 
events are very similar to large flares but on a much smaller spatial scale. 
Seconds after the reconnection takes place, a cloud of plasma is blown 
out from the BP. Madjarska (2009) reported that 
although  the event appears more like a jet (although some expanding loops 
can also be distinguished) in X-ray images as observed in projection on 
the solar disk, the two additional view points from STEREO/SECCHI reveal 
that the phenomenon evolves as an expulsion of BP loops followed 
by a collimated flow along the quasi-open field lines of the expanded 
loops. The escaping plasma reaches temperatures of around 2~MK 
\citep[][Madjarska 2009]{2007PASJ...59S.751C}. 

\citet{2009SoPh..tmp..120N} found 5 out of 79 jets analysed from STEREO/SECCHI 
observations exhibiting a three part structure typically of coronal mass 
ejections (CMEs) - bright leading edge, a dark void and 
bright trailing edge (e.g. the prominence material). The authors named 
them micro-CMEs. The rest were called Eiffel tower-type jets, where the 
reconnection appears to happen on top of the loops and lambda-type jets 
with reconnection occurring in the jet foot-points. As most of the events 
studied here were seen in projection on the solar disk, this visual 
division into different groups is not possible. However, the visual examination 
of the phenomena analysed  here confirms that expulsion of BP loops 
 describes these features best, hence, we will  further refer to them as EBPLs. 
\begin{figure*}[!ht]
\vspace{4.6cm}
\includegraphics{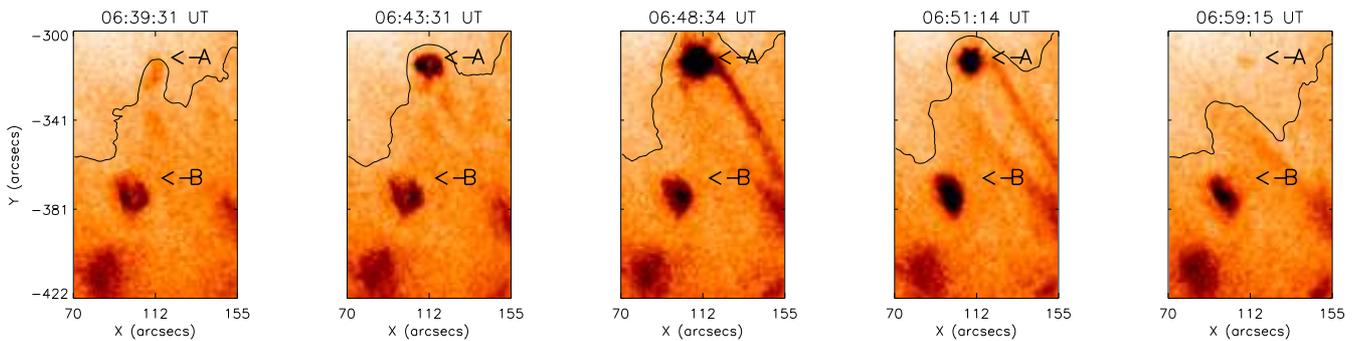}
\caption{An example of a typical jet-like happening on 2007 November 12 at the 
coronal hole boundaries. A refers to the jet while B refers to the BP.}
\label{fig5}
\end{figure*}


\begin{figure*}[!ht]
\vspace{3.2cm}
\includegraphics{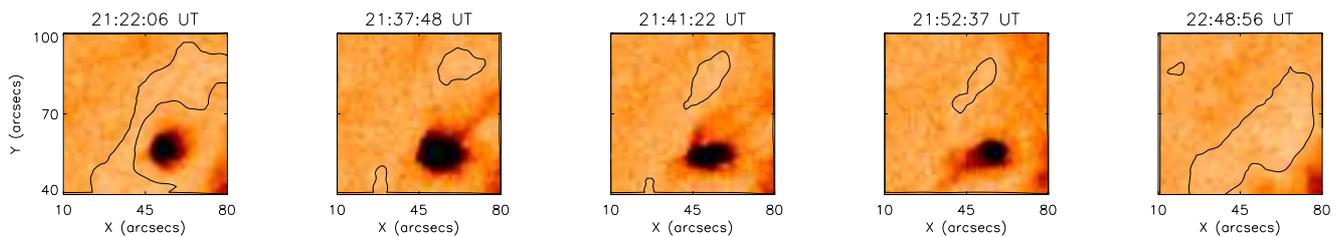}
\caption{A bright point which produced several jet-like events at the boundaries of a transient coronal hole on 2007 November 29.}
\label{fig6}
\end{figure*}


\begin{figure*}[!ht]
\vspace{3.2cm}
\includegraphics{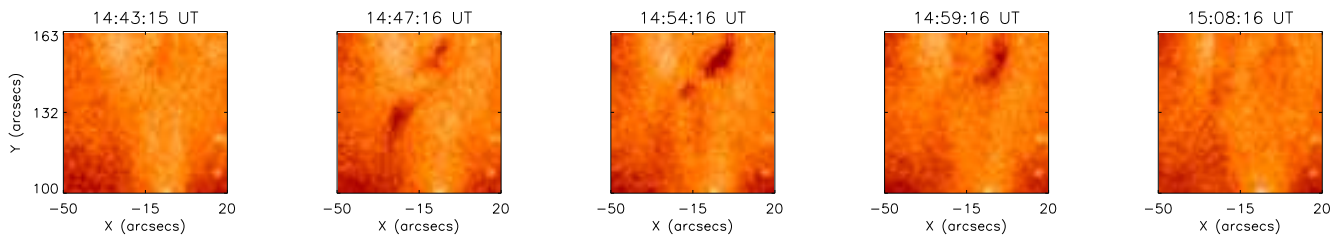}
\caption{An example of a brightening event identified in quiet Sun data  obtained on 2009 January 10.}
\label{fig7}
\end{figure*}
Fig.~\ref{fig5} presents an example of a typical EBPL event  
from a BP happening along an ECH boundary. The EBPL ends with the BP  
vanishing at X-ray temperatures triggering the coronal hole to expand.  
The plasma ejected from the BP seems to collide with structures on its way  
(propagating towards the QS region seen in projection on the solar disk) 
setting off a brightening in a pre-existing BP (denoted with B in 
Fig.~\ref{fig5}) with no obvious plasma outflows. 

We also found that the presence of a transient CH in the quiet Sun triggers 
the occurrence of EBPL-like phenomena, similar to the equatorial and 
polar CH ones. Yet again, the evolution of the BPs along the boundaries 
changed  the CHBs  (Fig.~\ref{fig6}).  Due to the smaller size of the TCH, 
these changes lead to a large expansion or contraction of the TCH, in some 
cases even a disappearance. In Fig.~\ref{fig7} we give a series of images 
which exhibit one of the quiet Sun events. In comparison to CHs, the 
events identified in  the quiet Sun images rarely show outflows 
(Table~\ref{table2}). Neither expanding loops nor collimated flows were 
distinguishable in the XRT images (i.e. no EBPL). 

This brings us to the numerical comparison of CHBRs and CHs with QS areas. We 
find an average of 75 brightening events per 100 $\times$ 100 arcsec$^2$ per day 
 for the coronal holes and boundaries and 20 events in the quiet Sun. Approximately 
70$\%$ of these brightenings in CHs and CHBRs showed plasma outflows while only 
30$\%$ of the brightenings seen  in the quiet Sun exhibit jet-like structures. 
Previous works indicated fewer events per day either because of the poorer spatial resolution of the 
instrument used or because of the visual identification methods. 

 The identified brightening events with no plasma outflows could either 
be driven by two sided loop reconnection \citep{1994xspy.conf...29S} between 
emerging fluxes and overlying coronal fields, in which the ejected plasma flow 
along the closed loop structures, or reconnection driven brightenings with 
plasma outflows at much lower temperatures. They can also represent flows in 
loop structures, perhaps triggered by reconnection shocks from the 
neighbourhood as seen in the example brightening event B in Fig.~\ref{fig5}. 
\citet{2007PASJ...59S.745S}  showed that a QS jet appears to be guided by the 
closed magnetic field lines (loops), unlike the jets in the CHs and CHBRs which 
are guided by the open magnetic field lines. This result immediately 
 raises the question whether the presence of open magnetic field lines is 
crucial for the generation of outflow phenomena. 

Comparison of the physical properties such as duration and size of the 
jet-like events in ECHs and polar CHs show no difference. A good 
correspondence was found between the duration and the size of the events, 
irrespective of their position. The larger events have longer duration 
of around 40 min and are mostly associated with pre-existing coronal 
bright points or at least with features becoming visible in X-rays just 
before the eruption. The smaller events have a shorter duration of around 
20 min (mostly with no pre-existing features at coronal temperatures).
However, for most events the actual duration from the moment the 
reconnection occurs (i.e. when the plasma is ejected) until the plasma 
outflow is no longer evident is usually between 10 and 15~{\rm min}.  
The amount of plasma ejected entirely depends on the magnetic 
energy available before the reconnection and it will, therefore, differ  
from event to event. Repetitive occurrence of jets in the same bright 
points are more common in CHs than in the QS. No periodicity was found, 
although a large number of BPs produced several jet-like events (from 2 
to 5 times) during the course of the observations until all the stored magnetic 
energy was exhausted and the bright point fully disappeared.  

 The proper motions of the outflows obtained from space-time plots are in
  the range of 100$-$500~\kms for most of the events. Because of the projection 
  effect of the jets with respect to the solar disk, the velocity we obtain gives the lower 
  boundary of the real velocity of the ejected plasma. Based 
on their plasma velocity we divide X-ray jets into two groups: (i) jets 
with pre-existing coronal structures (X-ray BPs) which have velocities 
$\approx$350~\kms\ or greater (i.e. in the range of the Alfv\'en velocity 
in the lower corona) and (ii) jets with no pre-existing structures at X-ray 
temperatures, showing velocities of around 150~\kms\ 
(i.e. close to the Alfv\'en velocity in the transition region). Stereo EUV images 
taken with the 171~\AA\ filter confirm the presence of a corresponding 
reconnection jet-like structures at transition region temperatures.  

\section{Conclusions}
\label{conclusions}

The present article confirms our findings from  paper I that the  evolution 
of loop structures known as coronal bright points is associated with the 
small-scale changes of CHBs.  We were able to identify the true nature of these 
changes which represent plasma outflows 
associated with the expansion of the bright point loop structures. The plasma 
trapped in the loop structures is consequently released along the ``quasi'' open 
magnetic field lines. These 
ejections appear to be triggered by magnetic reconnection, most probably  
the so-called interchange reconnection \citep{2004ApJ...612.1196W} between 
the closed magnetic field lines (BPs) and the open magnetic fields of coronal 
holes. The ejected plasma is guided and accelerated further away from the 
Sun by the open magnetic field lines with some jets reaching 
several solar radii \citep{2009SoPh..tmp..120N}. Contrary, in the quiet Sun 
the plasma ejected as a result of two or one sided loop reconnection, is 
contained in the corona by the closed magnetic field lines. 

 Tall and extended coronal loops are very rare in coronal holes 
\citep{2004SoPh..225..227W}, while closed (loop) magnetic structures of 
varying physical properties are ubiquitous in the quiet Sun corona as seen 
in EUV and X-ray observations. Coronal holes with predominant open magnetic 
fields and minority closed loop structures (BPs) are encompassed by these loop 
structures seen at transition region and coronal temperatures. 
\citet{2004ApJ...612.1171W}, \citet{2001ApJ...555..426W} and 
\citet{2005JGRA..11007109F} showed that the elemental abundance of  
trapped plasma is proportional to the confinement time of the plasma in 
loop structures. Newly emerged active region loop structures were found 
to have initial photospheric abundances (FIP bias $\approx$1--2)
 which increased with time, reaching $\approx$~5 in 1--3 days 
\citep{2001ApJ...555..426W}. \citet{1998Natur.394..152S} concluded that 
$\approx$1.5 days  is the reconfiguration time-scale for the super-granulation 
network magnetic fields  where coronal BPs have their foot-points rooted.  
This time period  is in the range of the confinement time-scale needed for 
the enhancement of the FIP bias. 

Coronal BPs electron densities were derived from CDS/SoHO in the temperature 
range 1.3--2$\times 10^6$~K by \citet{2005A&A...435.1169U}. The authors 
concluded that the bright points plasma have properties which are more 
similar to active region plasma rather than quiet Sun plasma, although BPs do 
not show the increase of electron density at temperature over 
Log$T_e$~$\sim$~6.2~K, observed in the core of active regions 
\citep{2005A&A...435.1169U}. These results were later confirmed from data taken 
with the Extreme-ultraviolet Imaging Spectrometer (EIS) for Log$T_e \sim$6.1 
and 6.2~K \citep {2008A&A...492..575P}. The BP lifetime in EUV were found to be 
on average 20~hrs in the EUV \citep{2001SoPh..198..347Z} and on average 8~hrs 
in X-rays with some BPs lasting up to 40~hrs \citep{1974ApJ...189L..93G}. The 
results of individually studied BPs by \citet{2004A&A...418..313U} give for two 
BPs a lifetime of 38 and 51~hrs detected in  Fe~{\sc xii}~195~\AA\  images from 
Extreme-ultraviolet Imaging Telescope (EIT) on-board SoHO and by P\'erez-Suarez 
(PhD thesis 2009) in five BPs: BP1 -- 48~hrs, BP2 more than 54~hrs, BP3 -- 
37~hrs, BP4 -- 45.2~hrs and BP5 -- 35 h (on the limb). The study by  
\citet{1974ApJ...189L..93G} on BPs lifetime in X-rays has not been updated so 
far using Hinode X-ray observations. The BPs properties given above strongly 
suggest that their plasma can become enriched on low FIP elements. 

 \citet{2008ApJ...682L.137R} concluded that X-ray jets are the precursors 
of polar plumes, and jets happening in pre-existing polar plumes enhance the 
brightness of the plume haze. Polar plumes are observed even at several 
solar radii \citep{1997SoPh..175..393D} and were found to contribute to the 
solar wind stream. They have been reported to occur even at low latitudes 
\citep{1995ApJ...446L..51W}.  Jets, associated with BPs, were also recently 
registered with the Large Angle and Spectrometric Coronagraph on-board SoHO 
\citep{2008SoPh..249...17W} and SECCHI/STEREO  \citep{2009SoPh..tmp..120N}. 
Hence, the  BP plasma cloud, which is ejected as a result of magnetic 
reconnection, will therefore, escape from the Sun having the plasma 
characteristics of the slow solar wind. We asked ourselves whether the plasma 
ejections we observe can possibly be a source of the fast solar wind? This 
possibility cannot be fully rejected, although it is a  fact that these jets 
happen sporadically rather than continuously, which is in contradiction with 
the nature of the fast solar wind.

Our specially designed observing programs provided us with spectroscopic 
co-observations from SUMER, CDS and EIS along with the XRT and SOT. In a  
follow up paper we will derive the physical properties such as velocity, 
density, temperature and others of a large number of events happening in the 
FOV of the spectrometers. 

\begin{acknowledgements} The authors thank ISSI, Bern for the support of 
the team ``Small-scale transient phenomena and their contribution to coronal 
heating''. Research at Armagh Observatory is grant-aided by the N.~Ireland 
Department of Culture, Arts and Leisure. We also thank STFC for support via 
grants ST/F001843/1 and PP/E002242/1. Hinode is a Japanese mission developed 
and launched by ISAS/JAXA, with NAOJ as domestic partner and NASA and STFC 
(UK) as international partners. It is operated by these agencies in 
co-operation with ESA and NSC (Norway). The STEREO/ SECCHI data used here are 
produced by an international consortium of the Naval Research Laboratory (USA), 
Lockheed Martin Solar and Astrophysics Lab (USA), NASA Goddard Space Flight 
Center (USA), Rutherford Appleton Laboratory (UK), University of Birmingham 
(UK), Max-Planck-Institut f\"{u}r Sonnensystemforschung (Germany), Centre Spatiale
de Li\`{e}ge (Belgium), Institut d'Optique Th\'{e}orique et Appliqu\'{e}e (France), and 
Institute Astrophysique Spatiale (France).

\end{acknowledgements}

\bibliographystyle{aa}

\begin{thebibliography}{34}
\expandafter\ifx\csname natexlab\endcsname\relax\def\natexlab#1{#1}\fi

\bibitem[{{Ahmad} \& {Webb}(1978)}]{1978SoPh...58..323A}
{Ahmad}, I.~A. \& {Webb}, D.~F. 1978, \solphys, 58, 323

\bibitem[{{Culhane} {et~al.}(2007){Culhane}, {Harra}, {Baker}, {van
  Driel-Gesztelyi}, {Sun}, {Doschek}, {Brooks}, {Lundquist}, {Kamio}, {Young},
  \& {Hansteen}}]{2007PASJ...59S.751C}
{Culhane}, L., {Harra}, L.~K., {Baker}, D., {et~al.} 2007, \pasj, 59, 751

\bibitem[{{Deforest} {et~al.}(1997){Deforest}, {Hoeksema}, {Gurman},
  {Thompson}, {Plunkett}, {Howard}, {Harrison}, \&
  {Hassler}}]{1997SoPh..175..393D}
{Deforest}, C.~E., {Hoeksema}, J.~T., {Gurman}, J.~B., {et~al.} 1997, \solphys,
  175, 393

\bibitem[{{DeForest} {et~al.}(2009){DeForest}, {Martens}, \&
  {Wills-Davey}}]{2009ApJ...690.1264D}
{DeForest}, C.~E., {Martens}, P.~C.~H., \& {Wills-Davey}, M.~J. 2009, \apj,
  690, 1264

\bibitem[{{DeForest} {et~al.}(2001){DeForest}, {Plunkett}, \&
  {Andrews}}]{2001ApJ...546..569D}
{DeForest}, C.~E., {Plunkett}, S.~P., \& {Andrews}, M.~D. 2001, \apj, 546, 569

\bibitem[{{Doyle} {et~al.}(2006){Doyle}, {Popescu}, \&
  {Taroyan}}]{2006A&A...446..327D}
{Doyle}, J.~G., {Popescu}, M.~D., \& {Taroyan}, Y. 2006, \aap, 446, 327

\bibitem[{{Feldman} {et~al.}(2005){Feldman}, {Landi}, \&
  {Schwadron}}]{2005JGRA..11007109F}
{Feldman}, U., {Landi}, E., \& {Schwadron}, N.~A. 2005, Journal of Geophysical
  Research (Space Physics), 110, 7109

\bibitem[{{Golub} {et~al.}(2007){Golub}, {Deluca}, {Austin}, {Bookbinder},
  {Caldwell}, {Cheimets}, {Cirtain}, {Cosmo}, {Reid}, {Sette}, {Weber},
  {Sakao}, {Kano}, {Shibasaki}, {Hara}, {Tsuneta}, {Kumagai}, {Tamura},
  {Shimojo}, {McCracken}, {Carpenter}, {Haight}, {Siler}, {Wright}, {Tucker},
  {Rutledge}, {Barbera}, {Peres}, \& {Varisco}}]{2007SoPh..243...63G}
{Golub}, L., {Deluca}, E., {Austin}, G., {et~al.} 2007, \solphys, 243, 63

\bibitem[{{Golub} {et~al.}(1974){Golub}, {Krieger}, {Silk}, {Timothy}, \&
  {Vaiana}}]{1974ApJ...189L..93G}
{Golub}, L., {Krieger}, A.~S., {Silk}, J.~K., {Timothy}, A.~F., \& {Vaiana},
  G.~S. 1974, \apjl, 189, L93+

\bibitem[{{Krieger} {et~al.}(1973){Krieger}, {Timothy}, \&
  {Roelof}}]{1973SoPh...29..505K}
{Krieger}, A.~S., {Timothy}, A.~F., \& {Roelof}, E.~C. 1973, \solphys, 29, 505

\bibitem[{{Madjarska} {et~al.}(2004){Madjarska}, {Doyle}, \& {van
  Driel-Gesztelyi}}]{2004ApJ...603L..57M}
{Madjarska}, M.~S., {Doyle}, J.~G., \& {van Driel-Gesztelyi}, L. 2004, \apjl,
  603, L57

\bibitem[{{Madjarska} \& {Wiegelmann}(2009)}]{2009arXiv0906.2556M}
{Madjarska}, M.~S. \& {Wiegelmann}, T. 2009, ArXiv e-prints

\bibitem[{{Moreno-Insertis} {et~al.}(2008){Moreno-Insertis}, {Galsgaard}, \&
  {Ugarte-Urra}}]{2008ApJ...673L.211M}
{Moreno-Insertis}, F., {Galsgaard}, K., \& {Ugarte-Urra}, I. 2008, \apjl, 673,
  L211

\bibitem[{{Nistic{\`o}} {et~al.}(2009){Nistic{\`o}}, {Bothmer}, {Patsourakos},
  \& {Zimbardo}}]{2009SoPh..tmp..120N}
{Nistic{\`o}}, G., {Bothmer}, V., {Patsourakos}, S., \& {Zimbardo}, G. 2009,
  \solphys, 120

\bibitem[{{P{\'e}rez-Su{\'a}rez} {et~al.}(2008){P{\'e}rez-Su{\'a}rez},
  {Maclean}, {Doyle}, \& {Madjarska}}]{2008A&A...492..575P}
{P{\'e}rez-Su{\'a}rez}, D., {Maclean}, R.~C., {Doyle}, J.~G., \& {Madjarska},
  M.~S. 2008, \aap, 492, 575

\bibitem[{{Raouafi} {et~al.}(2008){Raouafi}, {Petrie}, {Norton}, {Henney}, \&
  {Solanki}}]{2008ApJ...682L.137R}
{Raouafi}, N.-E., {Petrie}, G.~J.~D., {Norton}, A.~A., {Henney}, C.~J., \&
  {Solanki}, S.~K. 2008, \apjl, 682, L137

\bibitem[{{Saito}(1965)}]{1965PASJ...17....1S}
{Saito}, K. 1965, \pasj, 17, 1

\bibitem[{{Savcheva} {et~al.}(2007){Savcheva}, {Cirtain}, {Deluca},
  {Lundquist}, {Golub}, {Weber}, {Shimojo}, {Shibasaki}, {Sakao}, {Narukage},
  {Tsuneta}, \& {Kano}}]{2007PASJ...59S.771S}
{Savcheva}, A., {Cirtain}, J., {Deluca}, E.~E., {et~al.} 2007, \pasj, 59, 771

\bibitem[{{Schrijver} {et~al.}(1998){Schrijver}, {Title}, {Harvey}, {Sheeley},
  {Wang}, {van den Oord}, {Shine}, {Tarbell}, \&
  {Hurlburt}}]{1998Natur.394..152S}
{Schrijver}, C.~J., {Title}, A.~M., {Harvey}, K.~L., {et~al.} 1998, \nat, 394,
  152

\bibitem[{{Shibata} {et~al.}(1992){Shibata}, {Ishido}, {Acton}, {Strong},
  {Hirayama}, {Uchida}, {McAllister}, {Matsumoto}, {Tsuneta}, {Shimizu},
  {Hara}, {Sakurai}, {Ichimoto}, {Nishino}, \& {Ogawara}}]{1992PASJ...44L.173S}
{Shibata}, K., {Ishido}, Y., {Acton}, L.~W., {et~al.} 1992, \pasj, 44, L173

\bibitem[{{Shibata} {et~al.}(1994){Shibata}, {Nitta}, {Matsumoto}, {Tajima},
  {Yokoyama}, {Hirayama}, \& {Hudson}}]{1994xspy.conf...29S}
{Shibata}, K., {Nitta}, N., {Matsumoto}, R., {et~al.} 1994, in X-ray solar
  physics from Yohkoh, ed. Y.~{Uchida}, T.~{Watanabe}, K.~{Shibata}, \& H.~S.
  {Hudson}, 29--+

\bibitem[{{Shimojo} {et~al.}(1996){Shimojo}, {Hashimoto}, {Shibata},
  {Hirayama}, {Hudson}, \& {Acton}}]{1996PASJ...48..123S}
{Shimojo}, M., {Hashimoto}, S., {Shibata}, K., {et~al.} 1996, \pasj, 48, 123

\bibitem[{{Shimojo} {et~al.}(2007){Shimojo}, {Narukage}, {Kano}, {Sakao},
  {Tsuneta}, {Shibasaki}, {Cirtain}, {Lundquist}, {Reeves}, \&
  {Savcheva}}]{2007PASJ...59S.745S}
{Shimojo}, M., {Narukage}, N., {Kano}, R., {et~al.} 2007, \pasj, 59, 745

\bibitem[{{Ugarte-Urra} {et~al.}(2005){Ugarte-Urra}, {Doyle}, \& {Del
  Zanna}}]{2005A&A...435.1169U}
{Ugarte-Urra}, I., {Doyle}, J.~G., \& {Del Zanna}, G. 2005, \aap, 435, 1169

\bibitem[{{Ugarte-Urra} {et~al.}(2004){Ugarte-Urra}, {Doyle}, {Madjarska}, \&
  {O'Shea}}]{2004A&A...418..313U}
{Ugarte-Urra}, I., {Doyle}, J.~G., {Madjarska}, M.~S., \& {O'Shea}, E. 2004,
  \aap, 418, 313

\bibitem[{{von Steiger}(1996)}]{1996ASPC..109..491V}
{von Steiger}, R. 1996, in Astronomical Society of the Pacific Conference
  Series, Vol. 109, Cool Stars, Stellar Systems, and the Sun, ed.
  R.~{Pallavicini} \& A.~K. {Dupree}, 491--+

\bibitem[{{Wang} \& {Muglach}(2008)}]{2008SoPh..249...17W}
{Wang}, Y.-M. \& {Muglach}, K. 2008, \solphys, 249, 17

\bibitem[{{Wang} \& {Sheeley}(1995)}]{1995ApJ...446L..51W}
{Wang}, Y.-M. \& {Sheeley}, Jr., N.~R. 1995, \apjl, 446, L51+

\bibitem[{{Wang} \& {Sheeley}(2004)}]{2004ApJ...612.1196W}
{Wang}, Y.-M. \& {Sheeley}, Jr., N.~R. 2004, \apj, 612, 1196

\bibitem[{{Wang} {et~al.}(1998){Wang}, {Sheeley}, {Walters}, {Brueckner},
  {Howard}, {Michels}, {Lamy}, {Schwenn}, \& {Simnett}}]{1998ApJ...498L.165W}
{Wang}, Y.-M., {Sheeley}, Jr., N.~R., {Walters}, J.~H., {et~al.} 1998, \apjl,
  498, L165+

\bibitem[{{Widing} \& {Feldman}(2001)}]{2001ApJ...555..426W}
{Widing}, K.~G. \& {Feldman}, U. 2001, \apj, 555, 426

\bibitem[{{Wiegelmann} \& {Solanki}(2004)}]{2004SoPh..225..227W}
{Wiegelmann}, T. \& {Solanki}, S.~K. 2004, \solphys, 225, 227

\bibitem[{{Woo} {et~al.}(2004){Woo}, {Habbal}, \&
  {Feldman}}]{2004ApJ...612.1171W}
{Woo}, R., {Habbal}, S.~R., \& {Feldman}, U. 2004, \apj, 612, 1171

\bibitem[{{Zhang} {et~al.}(2001){Zhang}, {Kundu}, \&
  {White}}]{2001SoPh..198..347Z}
{Zhang}, J., {Kundu}, M.~R., \& {White}, S.~M. 2001, \solphys, 198, 347

\end{thebibliography}

\end{document}